\documentclass[aps,prd,preprint,floatfix]{revtex4} 

\usepackage{graphics}
\usepackage{graphicx}

\usepackage{latexsym}
\usepackage[cp866]{inputenc}
\usepackage[english]{babel}  

\input epsf

\begin{document}

\title{\boldmath Manifestations of the $a^0_0(980)-f_0(980)$ mixing
in $D^0\to K^0_S\pi^+\pi^-$\\ and $D^0\to K^0_S\eta\pi^0$ decays}
\author{N.~N.~Achasov and G.~N.~Shestakov}
\affiliation{Laboratory of Theoretical Physics, S.~L.~Sobolev
Institute for Mathematics, 630090 Novosibirsk, Russia}

\begin{abstract}
Possible manifestations of the isospin-breaking mixing of the
$a^0_0(980)$ and $f_0(980)$ resonances are analyzed in the $D^0\to
K^0_S \pi^+\pi^-$ and $D^0\to K^0_S\eta\pi^0$ decays. It is shown
that the $a^0_0(980)-f_0(980)$ mixing has the most influence on the
$\pi^+\pi^-$ mass spectrum in the decay $D^0\to K^0_S\pi^+\pi^-$.
Owing to the $a^0_0(980)-f_0(980)$ mixing, the $f_0(980)$ peak in
$D^0 \to K^0_Sf_0(980)\to K^0_S \pi^+\pi^-$ can experience
distortions comparable with its size. The effect essentially depends
on the relative phase between the amplitudes of the $D^0\to
K^0_Sf_0(980)$ and $D^0\to K^0_Sa^0_0(980)$ transitions.
\end{abstract}


\maketitle

Recently, we have examined the interference phenomena in the
$\eta\pi^0$ mass spectrum in the $D^+_s\to\eta\pi^0\pi^+$ decay
\cite{AS17} caused by the isospin-breaking $a^0_0(980)-f_0(980)$
mixing \cite{ADS79,AS16}. Their detection would help clarify the
production and decay mechanisms of the light scalars $f_0(980)$ and
$a^0_0(980)$ in weak three-body hadronic $D^+_s$ decays. A detailed
list of references concerning the $a^0_0(980)-f_0(980)$ mixing can
be found, for example, in Refs. \cite{AS16, AKS16}.

In Ref. \cite{AS17}, it has been noted that the investigations of
the $a^0_0(980)-f_0(980)$ mixing in three-body decays of the $D^0$
meson, such as $D^0\to K^0_S\pi^+\pi^-$, $D^0\to K^0_S\eta \pi^0$,
$D^0\to\bar K^0K^-K^+$, $D^0\to K^-K^+\pi^0 $, and $D^0\to
\pi^+\pi^-\pi^0$, are also promising and interesting. In the present
work, we analyze a possible manifestation of the $a^0_0(980)-f_0
(980)$ mixing in the decays $D^0\to K^0_S\pi^+\pi^-$ and $D^0\to
K^0_S\eta\pi^0$. We show that the $a^0_0(980)-f_0(980)$ mixing has
the most impact on the $\pi^+\pi^-$ mass spectrum in $D^0\to
K^0_S\pi^+\pi^-$. Owing to $a^0_0(980)-f_0(980)$ mixing, the
$f_0(980)$ peak in $D^0\to K^0_Sf_0(980)\to K^0_S \pi^+\pi^-$ can
experience distortions comparable with its size. The effect
essentially depends on the relative phase between the amplitudes of
the $D^0\to K^0_Sf_0(980)$ and $D^0\to K^0_Sa^0_0(980)$ transitions.

The $\pi^+\pi^-$ and $\eta\pi^0$ mass spectra in the $D^0\to K^0_S
\pi^+\pi^-$ and $D^0\to K^0_S\eta\pi^0$ decays caused by the $f_0(9
80)$ and $a^0_0(980)$ resonance contributions, respectively, and the
$a^0_0(980)-f_0(980)$ mixing have the form
\begin{equation}\label{Eq1}
\frac{dN_{\pi^+\pi^-}}{dm}=2mP_{K^0_S}(m)\rho_{\pi^+\pi^-}(m)
\frac{g^2_{f_0\pi^+\pi^-}}{16\pi} \left|\frac{C_1} {D_{f_0}
(m)}+\frac{e^{i\varphi}\,C_2\, \Pi_{a^0_0f_0}(m)} {D_{a^0_0}
(m)D_{f_0}(m)-\Pi^2_{a^0_0f_0}} \right|^2\,,
\end{equation}
\begin{equation}\label{Eq2}
\frac{dN_{\eta\pi^0}}{dm}=2mP_{K^0_S}(m)\rho_{\eta\pi^0}(m)
\frac{g^2_{a^0_0\eta\pi^0}}{16\pi} \left|\frac{e^{i\varphi}
\,C_2}{D_{a^0_0}(m)}+\frac{C_1\, \Pi_{a^0_0f_0}(m)}{D_{a^0_0}
(m)D_{f_0}(m)-\Pi^2_{a^0_0f_0}} \right|^2\,,
\end{equation}
where $m$ is the invariant mass of the $\pi^+\pi^-$ or $\eta\pi^0 $
system, and $\varphi$ is the relative phase between the amplitudes
of the $D^0\to K^0_Sf_0(980)$ and $D^0\to K^0_Sa^0_0(980)$
transitions. The kinematic factors
$P_{K^0_S}(m)=[m^4_{D^0}-2m^2_{D^0}(m^2_{K^0}+m^2)
+(m^2_{K^0}-m^2)^2]^{1/2}/(2m_{D^0})$, $\rho_{\pi^+\pi^-}(m)=(1
-4m^2_\pi/m^2)^{1/2}$, and $\rho_{\eta\pi^0}(m)=[1-2(m^2_\eta+
m^2_\pi)/m^2 +(m^2_\eta-m^2_\pi)^2/m^4]^{1/2}$; $m_\pi=m_\pi^+=
m_\pi^0=0.135$ GeV. Formulas for the inverse propagators
$D_{f_0}(m)$ and $D_{a^0_0}(m)$ of the $f_0(980)$ and $a^0_0(980)$
resonances and the amplitude $\Pi_{a^0_0 f_0}(m)$ describing the
$a^0_0(980)\to(K^+K^-+K^0\bar K^0)\to f_0(980)$ transition, together
with the values of coupling constants $g_{f_0ab}$, $ab=(\pi^+\pi^-,
\,\pi^0\pi^0,\,K^+K^-,\,K^0 \bar K^0,\,\eta\eta)$, and $g_{a^0_0ab}
$, $ab=(\eta\pi^0,\,K^+K^-,\,K^0\bar K^0,\,\eta'\pi^0)$, which we
use here, have been written in detail in Refs. \cite{AKS16,AS17}.
The values of $C_1=0.047$ GeV$^{-1/2}$ and $C_2=0.095$ GeV$^{-1/2}$
are fixed taking into account the CLEO Collaboration results
\cite{CLEO02,CLEO04}, together with the Particle Data Group
information \cite{PDG16}, and the relations
\begin{eqnarray}\label{Eq3}
BR(D^0\to K^0_Sf_0(980)\to K^0_S\pi^+\pi^-)
=\left(1.23^{+0.40}_{-0.24}\right)\cdot10^{-3}\nonumber\\
=\int\limits_{2m_\pi^+}^{m_{D^0}-m_{K^0}}
2mP_{K^0_S}(m)\rho_{\pi^+\pi^-}(m)\frac{g^2_{f_0\pi^+\pi^-}}{16\pi}
\left|\frac{C_1}{D_{f_0}(m)}\right|^2dm\,,
\end{eqnarray}
\begin{eqnarray}\label{Eq4}
BR(D^0\to K^0_Sa^0_0(980)\to K^0_S\eta\pi^0)
=\left(6.6\pm2.0\right)\cdot10^{-3}\nonumber\\
=\int\limits_{m_\eta+m_{\pi^0}}^{m_{D^0}-m_{K^0}}
2mP_{K^0_S}(m)\rho_{\eta\pi^0}(m)\frac{g^2_{a^0_0\eta\pi^0}}{16\pi}
\left|\frac{C_2}{D_{a^0_0}(m)}\right|^2dm\,.
\end{eqnarray}

The current experimental situation with the decays under
consideration will be discussed below. Here we only note that the
essential numerical input in Eqs. (\ref{Eq3}) and (\ref{Eq4}) is
based now on a very limited experimental statistics using the isobar
model of the decay amplitudes, and more precise information would be
very desirable. Unfortunately, in more recent analyses of the
$BABAR$ \cite{BaBar08} and Belle \cite{Belle14} Collaborations (see
also Ref. \cite{BaBar10}), the $f_0(980)$ contribution is not
separately treated in the fit to the data on the $D^0\to
K^0_S\pi^+\pi^-$ decay. Nevertheless, the $f_0(980)$ peak is clearly
visible in the $\pi^+\pi^-$ mass spectrum in this decay
\cite{BaBar08,Belle14}.

\begin{figure}
\centerline{\epsfysize=9.cm    
\epsfbox{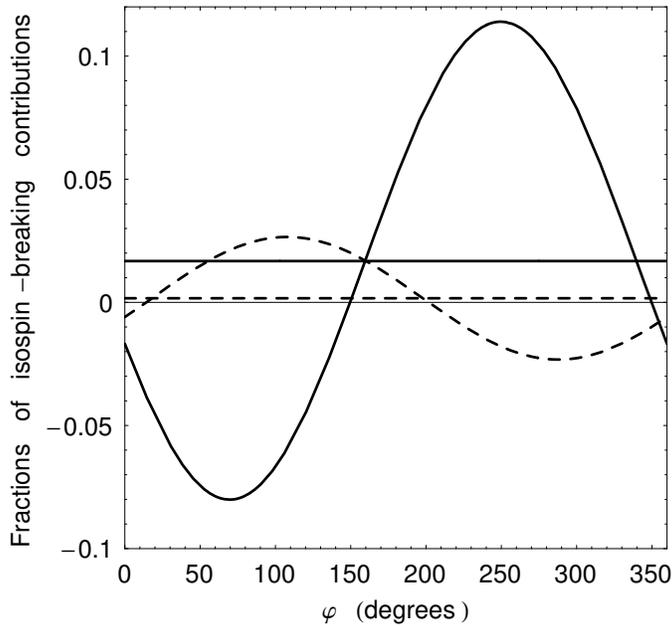}} 
\caption{The solid and dashed curves show the fractions of the
isospin-breaking contributions $\Delta BR(\pi^+\pi^- )$ and $\Delta
BR(\eta\pi^0)$ as functions of the relative phase $\varphi$,
respectively. The solid and dashed horizontal lines show the
incoherent contributions from the $a^0_0(980)-f_0(980)$ mixing in
the $\pi^+\pi^-$ and $\eta\pi^0$ channels, respectively.}
\label{Fig1}\end{figure}

The fractions of the $D^0\to K^0_S\pi^+\pi^-$ and $D^0\to
K^0_S\eta\pi^0$ branching ratios caused by the $a^0_0(980)-f_0(980)$
mixing are given by, respectively,
\begin{equation}\label{Eq5}
\Delta BR(\pi^+\pi^-)=\frac{1}{1.23\cdot10^{-3}}
\int\limits_{2m_\pi^+}^{m_{D^0}-m_{K^0}}
\frac{dN_{\pi^+\pi^-}}{dm}\,dm\ -1 
\end{equation} and
\begin{equation}\label{Eq6}
\Delta BR(\eta\pi^0)=\frac{1}{6.6\cdot10^{-3}}
\int\limits_{m_\eta+m_{\pi^0}}^{m_{D^0}-m_{K^0}}
\frac{dN_{\eta\pi^0}}{dm}\,dm\ -1\,.
\end{equation}
$\Delta BR(\pi^+\pi^-)$ and $\Delta BR(\eta\pi^0)$ as functions of
the relative phase $\varphi$ between the amplitudes of the $D^0\to
K^0_Sa^0_0(980)$ and $D^0\to K^0_Sf_0(980)$ decays are shown in Fig.
\ref{Fig1} by the solid and dashed curves, respectively. The solid
and dashed horizontal lines in Fig. \ref{Fig1} show the incoherent
contributions to $\Delta BR(\pi^+\pi^-)$ and $\Delta BR(\eta\pi^0)$
from the $a^0_0(980)-f_0 (980)$ mixing [i.e., the modules squared of
the second terms in the sums in Eqs. (\ref{Eq1}) and (\ref{Eq2})]
that are equal to $\approx1.7\%$ and $\approx0.17\%$, respectively.

\begin{figure}
\centerline{\epsfysize=17cm    
\epsfbox{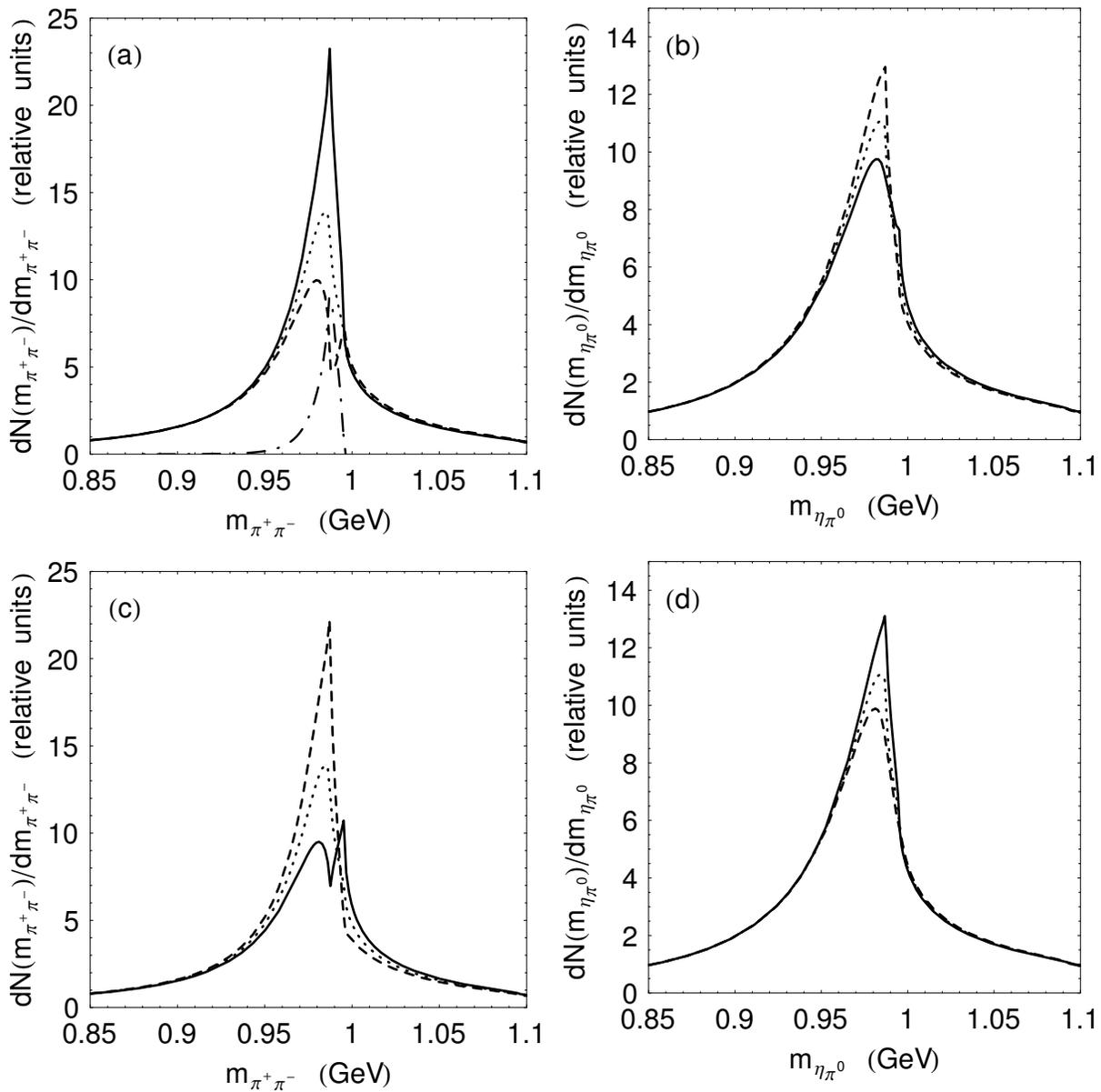}} 
\caption{The mass spectra $\pi^+\pi^-$ (a) and $\eta\pi^0$ (b) at
$\varphi=245^\circ$ (solid curves) and at $\varphi=70^\circ$ (dashed
curves); the dotted curves in (a) and (b) correspond to the mass
spectra without taking the $a^0_0(980)-f_0(980)$ mixing into
account, i.e., the $f_0(980)$ and $a^0_0(980)$ resonance
contributions, respectively. Plots (c) and (d) show the same as
plots (a) and (b) but for $\varphi=110^\circ$ (solid curves) and
$\varphi =290^\circ$ (dashed curves). As an example, the dot-dashed
curve in (a) shows the difference of the solid and dotted
curves.}\label{Fig2}
\end{figure}

As is seen from Fig. \ref{Fig1}, the maximal constructive
(destructive) interference between the contribution of the
$f_0(980)$ resonance and that of the $a^0_0(980)-f_0(980)$ mixing in
the $\pi^+\pi^-$ channel corresponds to the phase $\varphi\approx
245^\circ$ ($70^\circ$), while the maximal constructive
(destructive) interference between the contributions from the
$a^0_0(980)$ and the $a^0_0(980)-f_0(980)$ mixing in the $\eta\pi^0$
channel corresponds to the phase $\varphi\approx 110^\circ$
($290^\circ$). Figure \ref{Fig2} shows the mass spectra $dN_{
\pi^+\pi^-}/dm$ and $dN_{\eta\pi^0}/dm$ for these ultimate
interference patterns.

The fact that $BR(D^0\to K^0_Sa^0_0(980)\to K^0_S\eta\pi^0)$ is
about 5 times greater than $BR(D^0\to K^0_Sf_0(980)\to
K^0_S\pi^+\pi^-)$ [see Eqs. (\ref{Eq3}) and (\ref{Eq4})] leads to an
increase in the influence of the $a^0_0(980)-f_0(980)$ mixing on the
$\pi^+\pi^-$ mass spectrum in comparison with its influence in the
$\eta\pi^0$ channel. Owing to interference, the integral effect from
the $a^0_0(980)-f_0(980)$ mixing contribution can reach about $11\%$
(see Fig. \ref{Fig1}). This is very large for the isotopic
symmetry-breaking effect. As is seen from Fig. \ref{Fig2}, the
$a^0_0(980)-f_0(980)$ mixing can result in the essential distortions
of the $f_0(980)$ line shape in the $\pi^+\pi^-$ channel. For
example, it can lead to a narrowing of the $f_0(980)$ peak by about
1.5 times and to an increase of its height up to 60\% or even to the
formation of two peaks. The effect essentially depends on the
relative phase between the $D^0\to K^0_Sf_0(980)$ and $D^0\to
K^0_Sa^0_0(980)$ decay amplitudes. Certainly, a good mass resolution
and high statistics are required to detect phenomena of such a kind.

Let us briefly discuss the experimental situation \cite{PDG16,
CLEO04, BaBar05,CLEO02,BaBar08,Belle14,BaBar10}. There is only one
experiment on the $D^0\to K^0_S\eta\pi^0$ decay performed by the
CLEO Collaboration \cite{CLEO04}. The Dalitz analysis of 155
selected $D^0\to K^0_S\eta\pi^0$ candidates showed that the $D^0\to
K^0_S\eta\pi^0$ decay proceeds mainly via $K^0_S a^0_0(980)$ and
$K^{*0}(892)\eta$ intermediate states, the first of which is
dominant \cite{CLEO04}. Note that the $a^0_0(980)$ production has
been observed not only in the $D^0\to K^0_Sa^0_0 (980)\to
K^0_S\eta\pi^0$ decay [for which $BR(D^0\to K^0_Sa^0_0(980)\to
K^0_S\eta\pi^0)=(6.6\pm2.0)\cdot10^{-3}$ \cite{CLEO04,PDG16}] but
also in the channel $D^0\to K^0_S a^0_0(980)\to K^0_SK^+K^-$
\cite{BaBar05,BaBar10}. According to Ref. \cite{BaBar05}, $BR(D^0
\to K^0_Sa^0_0(980)\to K^0_SK^+ K^-)=(3.0\pm0.4)\cdot10^{-3}$, while
from Supplemental Material to Ref. \cite{BaBar10} (see Ref. [18] in
\cite{BaBar10}) it follows that the central value for $BR(D^0\to
K^0_Sa^0_0 (980)\to K^0_SK^+ K^-)\approx2.3\cdot10^{-3}$. The
relation of the above branching ratios is typical for the
$a^0_0(980)$ resonance in the $q^2\bar q^2$ model \cite{ADS81,AI89,
A98}, in which the $a_0(980)$ is strongly coupled with the $\eta\pi$
and $K\bar K$ channels.

In the CLEO \cite{CLEO02}, $BABAR$ \cite{BaBar08}, and Belle
\cite{Belle14} experiments 5299, 487\,000, and 1\,231\,731 $D^0\to
K^0_S\pi^+\pi^-$ candidates were selected, respectively. The $D^0\to
K^0_S\pi^+ \pi^-$ Dalitz distributions have a rich structure. Among
the possible intermediate states are such as $K^{*-}(982)\pi^+$,
$K^{*-}(1430)\pi^+$, $K^0_S\rho^0$, $K^0_Sf_0(980)$, $K^0_Sf_2(1270
)$, and $K^0_Sf_0(1370)$. According to the CLEO analysis
\cite{CLEO02}, the $K^0_S f_0(980)\to K^0_S\pi^+\pi^-$ fraction
constitutes of $(4.3^{+1.4}_{-0.8})\%$. The $K^0_S$ production
together with the $\pi^+\pi^-$ system in the $S$ wave,
$K^0_S(\pi^+\pi^-)_S$, constitutes about 12\% of the total branching
ratio of the $D^0$ decay into $K^0_S\pi^+\pi^-$
\cite{BaBar08,Belle14}. The $\pi^+\pi^-$ mass spectrum in the 1 GeV
region was scanned in the $BABAR$ \cite{BaBar08} and Belle
\cite{Belle14} experiments with an $\approx\,$5-MeV-wide step. It is
interesting to note that the whole visible $f_0(980)$ peak contains
only 6--7 points; i.e., its width is less than 25 MeV. Such a
narrowness of the $f_0(980)$ peak can be related to the effect of
the $a^0_0(980)-f_0(980)$ mixing. Of course, it is impossible to
completely eliminate the effect of interference of the $f_0(980)$
contribution with the background from other intermediate states.
There is hope that further investigations will clarify this issue.
The most clear information about the $D^0\to K^0_S f_0(980)\to K^0_S
\pi^+\pi^-$ decay channel, as well as about the possible role of
background contributions, is given by the distribution of $D^0\to
K^0_S\pi^+\pi^-$ events on the ($m^2_{\pi^+\pi^-},\, m^2_{K^0_S
\pi^\pm}$)-Dalitz plot.\\

The present work is partially supported by the Russian Foundation
for Basic Research Grant No. 16-02-00065 and the Presidium of the
Russian Academy of Sciences Project No. 0314-2015-0011.


\end{document}